\documentclass[12pt,a4paper]{article}
\usepackage{amsmath,amsthm,amsfonts,amssymb,alltt}
\usepackage{graphics}
\usepackage{graphicx}
\usepackage{epsfig}
\usepackage{color}
\usepackage{longtable}

\newtheorem{theorem}{Theorem}[section]

\theoremstyle{definition}

\theoremstyle{remark}
\newtheorem{remark}[theorem]{Remark}

\numberwithin{equation}{section}

\usepackage[section]{placeins}

\usepackage{bm}
\usepackage{mathtools}

\begin{document}

\author{Vladimir Mityushev\footnote{The corresponding author mityu@up.krakow.pl}, Wojciech Nawalaniec}

\title{Effective conductivity of a random suspension of highly conducting spherical particles}

\date{
Faculty of Mathematics, Physics and Technology, 
\\
Pedagogical University, ul.Podchorazych 2, Krakow, 30-084, Poland}
\maketitle

\abstract{
Randomly distributed non-overlapping perfectly conducting spheres are embedded in a conducting matrix  with the concentration of inclusions $f$. Jeffrey (1973) suggested an analytical formula valid up to $O(f^3)$ for macroscopically isotropic random composites. A conditionally convergent sum arose in the spatial averaging. In the present paper, we apply a method of functional equations to random composites and correct Jeffrey's formula. The main revision concerns the proper investigation of the conditionally convergent sum and correction the $f^2$-term. A new model of symbolic computations is developed in order to compute the effective conductivity tensor. The corresponding algorithm is realized up to $O(f^{\frac{10}3})$. The obtained formulae explicitly demonstrate the dependence of the effective conductivity tensor on the deterministic and probabilistic distributions of inclusions in the $f^2$-term, and in the $f^3$-term. This leads to the conclusion that some previous formulae presented as universal, i.e., valid for all random composites, may be actually applied only to dilute or to special composites when interaction between inclusions do not matter. 
}

Keywords: random composites; spherical inclusions; effective conductivity; Eisenstein summation; triple periodic functions ; symbolic computations

\section{Introduction}
\label{sec:intr}
The effective properties of composites with non-overlapping spherical inclusions are of considerable interest in a number of applied investigations. 
Due to universality of mathematical modeling we refer to conductivity and transport phenomena which arise in different physical problems, e.g., electrical conductivity, dielectrics, magnetism, heat conduction, diffusion, flow in porous media, and antiplane elasticity \cite{Weber,Milton}. Analytical approximate formulae for the macroscopic properties of random media are important in the fundamental sciences. They qualitatively describe the difference between regular and irregular composites, demonstrate precisely the dependence of their properties on geometrical structure. Such a dependence has the crucial importance in applications to polymer based composites and to nanocomposites. 

Clear comprehension of the impact of random geometry onto the macroscopic properties is established in statistical mechanics \cite{Krauth, Math} and in the theory of bounds \cite{Milton,Torquato,Cherkaev}. Analytical formulae for the effective properties of composites are used in material sciences to create composites with desired properties. However, engineers frequently derive and use analytical formulae valid for a narrow class of geometric structures, for instance dilute. What is more, they declare their universality. This mythology is shortly criticized in \cite{GMN}. It is difficult to divide valuable and wrong analytical formulae because wrong formulae are not wrong in peculiar cases. We do not mention doubtful results in the present paper since such a mention has to be justified. An exeption concerns only the paper \cite{Jeffrey} for which such a justification is made. It is worth noting that without any doubt many wrong results are useful in investigations and lead to the proper one.

We now proceed to review the discussed results on the mathematical level. 
Regular structure are sufficiently well studied by a method of multipole expansions 
\cite{Rayleigh,McPhedran1, McPhedran2, McPhedran3,McPhedran4, McPhedran5}, 
by Fourier series \cite{Zuzovsky, Sangani}, and by triply periodic functions \cite{Berdichevsky}. Asymptotically equivalent expressions for the effective conductivity were discussed in \cite{Andrianov} and \cite[Chapter 8]{GMN}.

\begin{figure}[!h]
	\centering
	\includegraphics[clip, trim=0mm 0mm 0mm 0mm, width=0.65\textwidth]{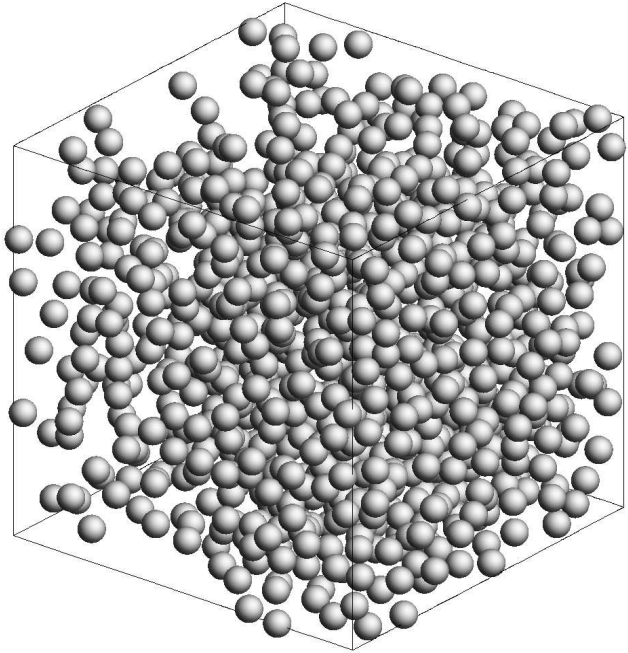}
	\caption{Randomly located non-overlapping spheres in the cubic cell. The method of generation is explained in Sec.\ref{sect:numRes}.}
\label{fig:sample}
\end{figure}

Various methods were developed to extend the above results to random composites, see an example of such a composite with $1000$ spheres per a periodicity cell is displayed in 
Fig.\ref{fig:sample}. We do not discuss here pure numerical methods useful for special geometries \cite{Zhang} and pay attention to analytical approximate formulae. 

Let equal spherical particles of conductivity $\lambda_1$ be embedded in a matrix of 
conductivity $\lambda$. Let $\beta = \frac{\lambda_1-\lambda}{\lambda_1+2\lambda}$ denote the contrast parameter. The first order approximation in $f$ for the effective conductivity is known as the Clausius-Mossotti approximation or Maxwell's formula \cite{Milton, Markov, Torquato} 
\begin{equation}
\label{eq:CMA}
\frac{\lambda_e}{\lambda}  = \frac{1+ 2 \beta f}{1-\beta f}+O(f^2)\;.
\end{equation}
In fluid mechanics, the effective viscosity of hard spherical particles can be found by the Einstein formula \cite{Markov, Torquato}
\begin{equation} 
\label{eq:Einstein}
\frac{\mu_e}{\mu} = 1+\frac{5}{2}f+O(f^2)\;,
\end{equation}         
where  $\mu$ stands for the viscosity of fluid.

Many attempts were applied to develop these formulae to high concentrations.  Theoretical difficulties were related to a conditionally convergent integral (sum) arisen in the spatial averaging. This integral for the effective conductivity of the regular cubic array was determined by Rayleigh \cite{Rayleigh}. Batchelor \cite{Batchelor} estimated a similar integral and the second order term of $\mu_e$ for random suspensions by rather physical intuition than a rigorous mathematical investigation. Jeffrey \cite{Jeffrey} modified Batchelor's approach to conductivity problem having taken into account two-sphere interactions and derived the following formula for macroscopically isotropic composites
\begin{equation}
\label{eq:Jeff}
\frac{\lambda_e}{\lambda} = 1+ 3\beta f+ 3\beta^2 f^2 + 3f^2 \beta^3 \left(\frac 34 +\frac 9{16} \frac{\lambda_1+2}{2\lambda_1+3}+ \ldots \right)\;.
\end{equation}
This formula was compared with others in \cite[p.493]{Torquato}. 

In the present paper, we revise Jeffrey's formula \eqref{eq:Jeff} and find that the $f^2$-term is equal to $3 f^2$ in the case $\beta=1$, i.e., for highly conducting inclusions. Jeffrey's formula \eqref{eq:Jeff} in this case gives $4.51 f^2$. We do not analyze where is the fallacy in \cite{Jeffrey}. We can say that it is certainly methodological, not a computational error, related to intuitive physical investigation of the conditionally convergent integral discussed in \cite{Jeffrey}. The critical review of the limit transition from finite to infinite number of inclusions in applications of self-consistent and cluster methods is presented in \cite{GMN,Mit2013,Mit2018,Mit2018a} for 2D conductivity problems. In particular, it was demonstrated that self-consistent and cluster methods with various intuitive  corrections are frequently applied within the accuracy $f$, but the obtained results are applied to high concentrations without any justification. 

The conductivity problem for a finite number of spheres in $\mathbb R^3$ was solved in \cite{MR} in terms of the 3D Poincar\'{e} series by a method of functional equations. This method can be treated as the alternating method of Schwrarz in the form of contrast or cluster expansions \cite{Mit2015}. The limit transition to infinite number of inclusions was performed in \cite[Chapter 8]{GMN} in the special 3D case when inclusions form the regular cubic lattice.

In the present paper, we develop the method of functional equations to the problem with $N$ spheres per unit cubic cell in Sec.\ref{secG}. First, the local field around a finite number $n$ of highly conducting spheres arbitrarily located without mutual intersections is exactly written in Sec.\ref{chap1:subsec3D-D}. The averaged local field is calculated in Sec.\ref{subsec22}. This first part follows the scheme \cite[Chapter 8]{GMN}. However, we cannot fully address to \cite{GMN}, since it was ultimately applied only to the regular cubic lattice. Here, we modify the method in order to apply it to an arbitrary location of inclusions. The limit $n \to \infty$ of the averaged local field is calculated in  Sec.\ref{sub:eff} by means of the Eisenstein summation following Rayleigh \cite{Rayleigh}. 
The justification of this approach was given in \cite[Sec.2.4]{McPhedran1} by study of the shape-dependent sums. It was shown that the Eisenstein summation yielded the vanishing polarization charge on the exterior surface having tended to infinity (for details see \cite[Fig.2]{McPhedran1}). Though this justification concerned a regular cubic lattice it can be extended verbatim to random composites. Application of the Eisenstein summation to random composites yields an analytical formula for the effective conductivity tensor. 
Its components are explicitly written in the form \eqref{eq:3Def6}-\eqref{eq:3Def6i} up to $O(f^{\frac{10}3})$.

Numerical examples with $N=1000$ spheres per cell are presented in Sec.\ref{sect:numRes}. The randomness is considered by the straight-forward approach used in \cite{GMN}. 
It can be shortly outlined as follows. First, a deterministic problem with an arbitrary location of non-overlapping spheres is solved and the effective conductivity tensor is explicitly written. Since the centers of spheres $\mathbf a_k$ and their number per cell $N$ are symbolically presented in the final formulae, we may consider them as the random variable  $\{ \mathbf a_1,  \mathbf a_2, \ldots,  \mathbf a_N\}$ which obeys a prescribed joint probabilistic distribution. We perform $10$ numerical experiments with $N=1000$ and obtain practically the same numerical values for the coefficients in the powers of $f$ in the expansion of the effective conductivity tensor. 
Sec.\ref{sec:Appl} is devoted to application of the obtained formulae to a stir casting process used for the fabrication of composites. Appendix A contains computational formulae for the lattice sums, Appendix B a new model of the applied symbolic computations. 
 
\section{General formula for highly conducting spheres}
\label{secG}
Let the vectors $\boldsymbol{\omega}_1=(1,0,0)$, $\boldsymbol{\omega}_2=(0,1,0)$ and $\boldsymbol{\omega}_3=(0,0,1)$ form a cubic lattice. The fundamental periodicity cell (the $\mathbf 0$-cell) is the cube $\mathcal O=\{\mathbf x \in \mathbb R^3: -\frac 12<x_j<\frac 12 \;(j=1,2,3)\}$. Let the centers $\mathbf a_k$ of mutually disjoint balls $D_k=\{\mathbf x  \in \mathbb R^3:|\mathbf x- \mathbf a_k|<r_k\}$ ($k=1,2, \ldots, N$) lie in $\mathcal O$ and $D$ denote the complement of all the balls $|\mathbf x- \mathbf a_k|\leq r_k$ to  $\mathcal O$. One can consider the triply periodic set of balls $\{D_k + \sum_{i=1,2,3}k_i \boldsymbol{\omega}_i\}$ ($k_i \in \mathbb Z$) equivalent to $N$ balls in the triply periodic topology. 

Let $\mathbf n=(n_1,n_2,n_3)$ denote the unit outward normal vector to the sphere $\partial D_k$ and $\frac{\partial}{\partial \mathbf n}$ the corresponding normal derivative. The normal vector has the form
\begin{equation}
\label{eq:3D1}
\mathbf n(\mathbf x) = \frac{1}{r_k} (\mathbf x- \mathbf a_k),\quad \mathbf x \in \partial D_k \quad (k=1,2, \ldots, n)\;.
\end{equation}

We are looking for functions $u(\mathbf x)$ harmonic in $D$ and $u_k(\mathbf x)$ harmonic in $D_k$ ($k=1,2,\ldots,N$), respectively, and continuously differentiable in the closures of the considered domains with the conjugation (transmission) conditions
\begin{equation}
\label{eq:3Da1}
u=u_k, \quad \frac{\partial u}{\partial \mathbf n} = \lambda_1
\frac{\partial u_k}{\partial \mathbf n},\quad |\mathbf x- \mathbf a_k|= r_k, \; k=1,2,\ldots,N\;.
\end{equation} 
Equations \eqref{eq:3Da1} express the perfect contact between materials of conductivity $\lambda_1$ occupying the balls and the host of the normalized unit conductivity.
It is assumed that the functions $u(\mathbf x)$ and $u_k(\mathbf x)$ are quasi-periodic, namely,
\begin{equation}
\label{eq:3Da2}
[u]_1=1, \quad [u]_2=0, \quad [u]_3=0\;, 
\end{equation} 
where $[u]_j:=u(\mathbf x +\omega_j)-u(\mathbf x)$ stands for the jump of $u(\mathbf x)$ per cell along the axis $x_j$.

Instead of the normal derivative in \eqref{eq:3Da1} we will also consider the derivative $\frac{\partial}{\partial r}$ where $r=|\mathbf x- \mathbf a_k|$ is the radial local coordinate near $\partial D_k$. For a fixed $k$, we have 
\begin{equation}
\label{eq:3Da1r1}
\frac{\partial}{\partial \mathbf n}= 
\frac{x_1-a_{k1}}{r_k} \frac{\partial}{\partial x_1}+
\frac{x_2-a_{k2}}{r_k} \frac{\partial}{\partial x_2}+
\frac{x_3-a_{k3}}{r_k} \frac{\partial}{\partial x_3}
= \frac{\partial}{\partial r}, \quad r=r_k\;,
\end{equation}  
where $\mathbf a_k =(a_{k1},a_{k2},a_{k3})$. Then, \eqref{eq:3Da1} becomes
\begin{equation}
\label{eq:3Da1r}
u=u_k, \quad \frac{\partial u}{\partial r} = \lambda_1
\frac{\partial u_k}{\partial r},\quad r= r_k, \; k=1,2,\ldots,N\;.
\end{equation}   

The averaged flux $\langle\mathbf q \rangle$ can be calculated by application of the Ostrogradsky - Gauss formula in $\mathcal O$. The flux component $\langle q_i \rangle$ along the axis $x_i$ becomes 
\begin{align}
\label{eq:3Da3}
-\langle q_i \rangle = \int_D \frac{\partial u}{\partial x_i} d\mathbf x +
\lambda_1 \sum_{k=1}^N \int_{D_k} \frac{\partial u_k}{\partial x_i} d\mathbf x 
= \\ \nonumber
\int_{\partial \mathcal O} u\; n_i ds +
 \sum_{k=1}^N \int_{\partial D_k} \left(\lambda_1 u_k - u\right) n_i ds\;, 
\end{align}
where $\mathbf n=(n_1,n_2,n_3)$. 
The integral over $\partial \mathcal O = \partial D - \sum_{k=1}^n \partial D_k$ in \eqref{eq:3Da3} is equal to the Kronecker delta $\delta_{i1}$ because of \eqref{eq:3Da2}. 
Using the first relation \eqref{eq:3Da1} and again the Ostrogradsky-Gauss formula, we obtain
\begin{equation}
\label{eq:3Da5}
-\langle q_i \rangle = \delta_{i1}+ (\lambda_1-1)\sum_{k=1}^N 
\int_{D_k} \frac{\partial u_k}{\partial x_i} d\mathbf x\;.
\end{equation} 
The mean value theorem for harmonic functions \cite[p.294]{TS} yields
\begin{equation}
\label{eq:3Da6}
-\langle q_i \rangle = \delta_{i1}+ (\lambda_1-1)\sum_{k=1}^N \frac 43 \pi r_k^3\;
 \frac{\partial u_k}{\partial x_i} (\mathbf a_k)\;.
\end{equation} 
Let $\Lambda=\{\lambda_{ij}\}$ denote the effective conductivity tensor defined by the relation  $\mathbf q=-\Lambda \langle \overline{\nabla u} \rangle$. Here, $\overline{\nabla u}=([u]_1,[u]_2,[u]_3)$ stands for the jump vector of the potential $u$ per cell. The normalized jump conditions  \eqref{eq:3Da2} yield $\lambda_{ij}=-\langle q_i \rangle$, hence, \eqref{eq:3Da6} becomes
\begin{equation}
\label{eq:3Da67}
\lambda_{ij} = \delta_{i1}+ (\lambda_1-1)\sum_{k=1}^N \frac 43 \pi r_k^3\;
 \frac{\partial u_k}{\partial x_i} (\mathbf a_k)\;.
\end{equation} 
The formula \eqref{eq:3Da67} is derived for $j=1$, i.e., for the external flux applied in the $x_1$-direction, see \eqref{eq:3Da2}, and can be established analogously for $j=2,3$. Therefore, the functions $u_k(\mathbf x)$ from \eqref{eq:3Da7} depend also on $j$.

In the case of equal radii we arrive at the formula
\begin{equation}
\label{eq:3Da7}
\lambda_{ij}= \delta_{ij}+ (\lambda_1-1)\; \frac fN \sum_{k=1}^N 
 \frac{\partial u_k}{\partial x_i} (\mathbf a_k) \quad (i,j=1,2,3)\;,
\end{equation} 
where $f$ denote the concentration of inclusions. 

Below, we study the case of highly conducting inclusions when $\lambda_1\gg 1$ and 
\begin{equation}
\label{eq:3Da8}
u= u^{(0)} + \frac 1{\lambda_1} u^{(1)} + \frac 1{\lambda_1^2} u^{(2)} + 
\ldots, \; u_k= u^{(0)}_k + \frac 1{\lambda_1} u^{(1)}_k + \frac 1{\lambda_1^2} u^{(2)}_k +\ldots \;.
\end{equation} 
Substitute \eqref{eq:3Da8} into \eqref{eq:3Da1} and take the terms up to $O(\lambda_1^{-1})$. The zero-th coefficient yields the boundary value problem
\begin{align}
\label{eq:3Da9a}
u^{(0)}=c_k, 
\quad |\mathbf x- \mathbf a_k|= r_k \; (k=1,2,\ldots,N)\;.
\end{align} 
Equation \eqref{eq:3Da9a} can be considered as the modified Dirichlet problem for the exterior of spheres with the undetermined constants $c_k$ on the boundary. It follows from \cite[Chapter 8]{GMN} that the leading part of the effective conductivity tensor is expressed by means of the function $u^{(0)}(\mathbf x)$ by formula
\begin{equation}
\label{eq:3Da16}
\lambda_{ij}= \delta_{ij}+ \sum_{k=1}^N  \frac 43 \pi r_k^3\; \int_{\partial D_k}(x_i-a_{ki}) \frac{\partial u^{(0)}}{\partial r} ds+O(\lambda_1^{-1})\;.
\end{equation}

\section{Modified Dirichlet problem}
\label{chap1:subsec3D-D}
In the present section, we solve the modified Dirichlet problem \eqref{eq:3Da9a} by the method of functional equations. This method was proposed in \cite{MR} for a finite number of inclusions $n$ in $\mathbb R^3$. The method was extended to triple periodic problems by the limit $n\to \infty$ in \cite[Chapter 8]{GMN} for $N=1$. 

First, we shortly present the functional equations and their solution for a finite number of inclusions. The rigorous justification of the method of functional equations can be found in \cite{GMN,MR}. It is worth noting that a system of linear algebraic equations on the undetermined constants $c_k$ was constructed but not solved in \cite{GMN}. Next, we apply an asymptotic method to solve this system for equal radii and find the undetermined constants in the form of series. The obtained explicit formulae are used in the next sections to determine the effective conductivity tensor.  
 
In the present section, we consider a problem for a finite domain $D(n)$ with $n$ balls in $\mathbb R^3$ and construct its solution $u(\mathbf x,n)$. The required triple periodic function can be constructed by the limit transition $u^{(0)}(\mathbf x) = \lim_{n \to \infty} u(\mathbf x,n)$. It will be convenient to introduce auxiliary functions $u_k(\mathbf x,n)$ harmonic in $D_k$ and express the effective conductivity in terms of $\frac{\partial u_k}{\partial x_i} (\mathbf a_k,\infty)$. Formally, $u_k(\mathbf x,n)$ differs from the potential $u_k(\mathbf x)$ in $D_k$ for the general problem \eqref{eq:3Da1} with finite $\lambda_1$. The new function $u_k(\mathbf x,n)$ can be considered as the potential in $D_k$ satisfying the special interface condition \eqref{eq:3D2-15} discussed below.
We write $D$ and $u(\mathbf x)$ instead of  $D(n)$ and $u(\mathbf x,n)$ for shortness.  

Consider mutually disjoint balls $D_k=\{\mathbf x  \in \mathbb R^3:|\mathbf x- \mathbf a_k|<r_k\}$ ($k=1,2, \ldots, n$) and the domain $\dot{D}= \mathbb R^3 \backslash \cup_{k=1}^n (D_k \cup \partial D_k)$. It is convenient to add the infinite point and introduce the domain $D=\dot{D}\cup \{\infty\}$ lying in the one-point compactification of $\mathbb R^3$.
We find a function $u(\mathbf x)$ harmonic in $\dot{D}$ and continuously differentiable in $\dot{D}\cup_{k=1}^n \partial D_k$ with the boundary conditions
\begin{equation}
\label{eq:3D2}
u=c_k, \quad  \mathbf x \in \partial D_k \quad (k=1,2, \ldots, n)\;.
\end{equation}
Here, $c_k$ are undetermined constants which should be found during solution to the boundary value problem. Let the external flux at infinity be parallel to the $x_j$-axis, hence,
\begin{equation}
\label{eq:3D21}
u(\mathbf x)-x_j  \; \mbox{tends to 0, as} \;  |\mathbf x| \to \infty\;.
\end{equation} 
We have
\begin{equation}
\label{eq:3D212}
\int_{\partial D_k} \frac{\partial u}{\partial \mathbf n} ds =0 
\Longleftrightarrow
\int_{\partial D_k} \frac{\partial u}{\partial r} ds =0 
\quad (k=1,2, \ldots, n)\;.
\end{equation} 

The inversion with respect to a sphere $\partial D_k$ is introduced by formula
\begin{equation}
\label{eq:3D3}
\mathbf x_{(k)}^* = \frac{r_k^2}{r^2} (\mathbf x- \mathbf a_k) + \mathbf a_k,
\end{equation}
where $r=|\mathbf x- \mathbf a_k|$ denotes the local spherical coordinate near the point $\mathbf x= \mathbf a_k$\;.
The Kelvin transform with respect to $\partial D_k$ has the form \cite{Weber,TS} 
\begin{equation}
\label{eq:3D34}
\mathcal K_k w(\mathbf x) = \frac{r_k}{r} w(\mathbf x_{(k)}^*)\;.
\end{equation}
If a function $w(\mathbf x)$ is harmonic in $|\mathbf x- \mathbf a_k|<r_k$, the function $\mathcal K_k w(\mathbf x)$ is harmonic in $|\mathbf x- \mathbf a_k|>r_k$ and vanishes at infinity.

The method of functional equations was developed in \cite{MR,GMN} to solve the modified Dirichlet problem \eqref{eq:3D2}. First, auxiliary functions $u_k(\mathbf x)$ harmonic in the balls $|\mathbf x- \mathbf a_k| < r_k$, respectively, and continuously differentiable in their closures, satisfying  the boundary conditions 
\begin{equation}
\label{eq:3D2-15}
\frac{\partial u}{\partial \mathbf n}=
\frac{u_k}{r_0}+2\frac{\partial u_k}{\partial \mathbf n},\quad |\mathbf x- \mathbf a_k| = r_k, 
\;(k=1,2, \ldots, n)  
\end{equation}
were introduced. Second, it was proved that these functions satisfy the system of functional equations
\begin{equation}
\label{eq:3D37}
u_k(\mathbf x)=-\sum_{m \neq k} \frac{r_m}{|\mathbf x- \mathbf a_m|} u_m(\mathbf x_{(m)}^*)+x_j-c_k, \; |\mathbf x- \mathbf a_k| \leq r_k, 
\;k=1,2, \ldots, n  
\end{equation}
and the conditions 
\begin{equation}
\label{eq:3D212c}
 \int_{\partial D_k}  u_k  ds = 0 \quad (k=1,2, \ldots, n)\;.
\end{equation} 
We now fix the constants $c_k$ in \eqref{eq:3D37} and return to their determination at the end of this section.
The functional equations \eqref{eq:3D37} are solved by the method of successive approximations uniformly convergent in the union of the balls $|\mathbf x- \mathbf a_k| \leq r_k$, see \cite{GMN} for details. 

After solution to the system \eqref{eq:3D37} the solution of the problem  \eqref{eq:3D21} is given up to undetermined constants by formula
\begin{equation}
\label{eq:3D38}
u(\mathbf x) =-\sum_{m=1}^n \frac{r_m}{|\mathbf x- \mathbf a_m|} u_m(\mathbf x_{(m)}^*)+x_j,  \quad \mathbf x \in \dot{D}\;.
\end{equation}

Introduce the space $\mathcal C(D_k)$ of functions harmonic in all $D_k$ and continuous in their closures endowed with the norm $\|h\|_k = \max_{\mathbf x \in \partial D_k} |h(\mathbf x)|$.
Let $\mathcal C=\mathcal C\left( \cup_{k=1}^n D_k \right)$ denote the space of functions harmonic in all $D_k$ and continuous in their closures. The norm in $\mathcal C$ has the form $\|h\| =\max_k \|h\|_k = \max_k \max_{\mathbf x \in \partial D_k} |h(\mathbf x)|$. The system of functional equations \eqref{eq:3D37} can be considered as an equation in the space $\mathcal C$ on a function equal to $u_k(\mathbf x)$ in each closed ball $D_k \cup \partial D_k$.

Let $k_s$ run over $1,2, \ldots, n$.
Consider the sequence of inversions with respect to the spheres $\partial D_{k_1},\partial D_{k_2}, \ldots, \partial D_{k_m}$ determined by the recurrence formula
\begin{equation}
\label{eq:3DS}
x_{(k_m k_{m-1}\ldots k_1)}^* :=\left(x_{(k_{m-1}k_{m-2}\ldots k_1)}^* \right)_{k_m}^*\;.
\end{equation}
It is supposed that no equal neighbor numbers in the sequence $k_1, k_2, \ldots, k_m$. The transformations \eqref{eq:3DS} for $m=1,2,\ldots$ with the identity map form the Schottky group $\mathcal S$ of maps acting in $\mathbb R^3$, see the 2D theory \cite{Mit2011}. 

Introduce the function $h(\mathbf x)$ in the space $\mathcal C$ defined by equations $h(\mathbf x)=x_j-c_k$ in $|\mathbf x- \mathbf a_k| \leq r_k$ ($k=1,2, \ldots, n$).
Straight-forward application of the successive approximations to \eqref{eq:3D37} gives the exact formula  
\begin{equation}
\label{eq:Poincare1}
u_k(\mathbf x) = (P h) (\mathbf x), \quad \mathbf x \in D_k \cup \partial D_k, 
\end{equation}
where the operator $P$ acts in the space $\mathcal C$ and has the form $P=P_k$ in $D_k \cup \partial D_k$; the operator $P_k$ is defined in terms of the series
\begin{align}
\label{eq:Poin1}
(P_k h) (\mathbf x)  = h(\mathbf x)-
\\
\notag
\sum_{m\neq k} \frac{r_m}{|\mathbf x- \mathbf a_m|} h(\mathbf x_{(m)}^*)+ 
\sum_{m\neq k} \sum_{k_1\neq k} \frac{r_m}{|\mathbf x- \mathbf a_m|} 
\frac{r_{k_1}}{|\mathbf x_{(m)}^*- \mathbf a_{k_1}|} h(\mathbf x_{(k_1m)}^*) -  
\\
\notag
\sum_{m\neq k} \sum_{\substack{k_1\neq m\\k_2\neq k_1}}\frac{r_m}{|\mathbf x- \mathbf a_m|} 
\frac{r_{k_1}}{|\mathbf x_{(m)}^*- \mathbf a_{k_1}|} \frac{r_{k_2}}{|\mathbf x_{(k_1m)}^*- \mathbf a_{k_2}|} h(\mathbf x_{(k_2k_1m)}^*)+ \ldots \;,
\end{align}
where for instance $\sum_{\substack{k_1\neq m\\k_2\neq k_1}}:=\sum_{k_1\neq m} \sum_{k_2\neq k_1}$. 
Every sum $\sum_{k_{s}\neq k_{s-1}}$ contains terms with $k_{s} = 1,2, \ldots, n$ except $k_{s} = k_{s-1}$. Following \cite[Chapter 4]{MiR} and \cite{MR} one can prove compactness of $P$ in $\mathcal C$. 

The uniqueness of solution established in \cite{MR,GMN} implies that the successive approximations can be applied separately to $h_1(\mathbf x)$ and to $h_2(\mathbf x)$ when $h(\mathbf x) = h_1(\mathbf x) + h_2(\mathbf x)$. The unique result will be the same.  Thus, $P$ can be applied separately to $x_j$ and to the piece-wise constant function $c(\mathbf x) = c_k$ where $x \in D_k$ ($k=1,2, \ldots, n$).  Then, the solution of the system \eqref{eq:3D37} can be written in the form
\begin{equation}
\label{eq:Poincare2}
u_k(\mathbf x)  = (P x_j) (\mathbf x) - (P c)(\mathbf x), \quad \mathbf x \in D_k \cup \partial D_k \; (k=1,2, \ldots, n)\;.
\end{equation}

Substitution of \eqref{eq:Poincare2} into \eqref{eq:3D38} yields 
\begin{equation}
\label{eq:Poin3}
u(\mathbf x)  =x_j+ (P_0 x_j) (\mathbf x) - (P_0 c)(\mathbf x),  \quad \mathbf x \in \dot{D}\;,
\end{equation} 
where the operator $P_0$ is introduced as follows
\begin{align}
\label{eq:PoiD}
(P_0 h)(\mathbf x) = 
-\sum_{k=1}^n \frac{r_k}{|\mathbf x- \mathbf a_k|} h(\mathbf x_{(k)}^*)
\\
\notag
+ \sum_{k=1}^n \sum_{k_1\neq k} \frac{r_k}{|\mathbf x- \mathbf a_k|} 
\frac{r_{k_1}}{|\mathbf x_{(k)}^*- \mathbf a_{k_1}|} h(\mathbf x_{(k_1k)}^*) -  
\\
\notag
\sum_{k=1}^n \sum_{\substack{k_1\neq k\\k_2\neq k_1}} \frac{r_k}{|\mathbf x- \mathbf a_k|} 
\frac{r_{k_1}}{|\mathbf x_{(k)}^*- \mathbf a_{k_1}|} \frac{r_{k_2}}{|\mathbf x_{(k_1k)}^*- \mathbf a_{k_2}|}  h(\mathbf x_{(k_2k_1k)}^*)
\end{align}
\begin{align}
\nonumber
+\sum_{k=1}^n \sum_{\substack{k_1\neq k\\k_2\neq k_1 \\ k_3\neq k_2}}
\frac{r_k}{|\mathbf x- \mathbf a_k|} 
\frac{r_{k_1}}{|\mathbf x_{(k)}^*- \mathbf a_{k_1}|} 
\frac{r_{k_2}}{|\mathbf x_{(k_1k)}^*- \mathbf a_{k_2}|} 
\frac{r_{k_3}}{|\mathbf x_{(k_2k_1k)}^*- \mathbf a_{k_3}|} h(\mathbf x_{(k_3k_2k_1k)}^*)
\\
\notag
+ \ldots,  \quad \mathbf x \in \dot{D}\;.
\end{align}

Consider a class of functions $\mathcal R$ harmonic in $\mathbb R^3$ except a finite set of isolated points located in $D$ where at most polynomial growth can take place. It follows from \eqref{eq:Poin1} that the operator  $P: \mathcal R \to \mathcal R$ is properly defined.  
The series  $(P_0h) (\mathbf x)$ for $h \in \mathcal R$  converges uniformly in every compact subset of  $\mathbf x \in \dot{D}\cup \partial D$. We call it by the 3D Poincar\'{e} $\theta$-series associated to the Schottky group $\mathcal S$ for the function $h(\mathbf x)$.  

One can see from \eqref{eq:Poin3} that $u(\mathbf x)$ contains the undetermined constants $c_k$ only in the part
\begin{align}
\label{eq:Poi1}
(P_0 c)(\mathbf x) = 
-\sum_{k=1}^n \frac{r_k}{|\mathbf x- \mathbf a_k|} c_k+ 
\sum_{k=1}^n \sum_{k_1\neq k} \frac{r_k}{|\mathbf x- \mathbf a_k|} 
\frac{r_{k_1}}{|\mathbf x_{(k)}^*- \mathbf a_{k_1}|} c_{k_1} -  
\\
\notag
\sum_{k=1}^n \sum_{\substack{k_1\neq k\\k_2\neq k_1}} \frac{r_k}{|\mathbf x- \mathbf a_k|} 
\frac{r_{k_1}}{|\mathbf x_{(k)}^*- \mathbf a_{k_1}|} \frac{r_{k_2}}{|\mathbf x_{(k_1k)}^*- \mathbf a_{k_2}|} c_{k_2}+
\end{align}
\begin{equation}
\nonumber
\sum_{k=1}^n \sum_{\substack{k_1\neq k\\k_2\neq k_1 \\ k_3\neq k_2}}
\frac{r_k}{|\mathbf x- \mathbf a_k|} 
\frac{r_{k_1}}{|\mathbf x_{(k)}^*- \mathbf a_{k_1}|} 
\frac{r_{k_2}}{|\mathbf x_{(k_1k)}^*- \mathbf a_{k_2}|} 
\frac{r_{k_3}}{|\mathbf x_{(k_2k_1k)}^*- \mathbf a_{k_3}|}c_{k_3}+
\ldots \;.
\end{equation}
The function $(P_0 c)(\mathbf x)$ is represented in the form $(P_0 c)(\mathbf x) = \sum_{m=1}^n c_m p_m(\mathbf x)$. Let $\delta_{km}$ denote the Kronecker symbol.  The functions $p_m(\mathbf x)$ can be obtained by Lemma 4.3 from \cite[p.152]{MiR} or by substitution  $c(\mathbf x) = \delta_{km}$, $\mathbf x \in D_k$ ($k=1,2,\ldots,n$) into \eqref{eq:Poi1} 
\begin{align}
\label{eq:Poi2}
p_m(\mathbf x) = 
 -\frac{r_m}{|\mathbf x- \mathbf a_m|} +
\sum_{k\neq m} \frac{r_k}{|\mathbf x- \mathbf a_k|} \frac{r_{m}}{|\mathbf x_{(k)}^*- \mathbf a_{m}|}   -  
\\
\notag
\sum_{k=1}^n \sum_{\substack{k_1\neq k,m}} \frac{r_k}{|\mathbf x- \mathbf a_k|} 
\frac{r_{k_1}}{|\mathbf x_{(k)}^*- \mathbf a_{k_1}|} \frac{r_{m}}{|\mathbf x_{(k_1k)}^*- \mathbf a_{m}|} +
\end{align}
\begin{align}
\notag
\sum_{k=1}^n \sum_{\substack{k_1\neq k\\k_2\neq k_1,m }}
\frac{r_k}{|\mathbf x- \mathbf a_k|} 
\frac{r_{k_1}}{|\mathbf x_{(k)}^*- \mathbf a_{k_1}|} 
\frac{r_{k_2}}{|\mathbf x_{(k_1k)}^*- \mathbf a_{k_2}|} 
\frac{r_{m}}{|\mathbf x_{(k_2k_1k)}^*- \mathbf a_{m}|}+
\ldots \;.
\end{align}

In order to find the undetermined constants $c_k$ $(k=1,2, \ldots, n)$ we use equations
\eqref{eq:3D212c}. The integral from \eqref{eq:3D212c} is calculated by the mean value theorem for harmonic functions \cite[p.294]{TS}. Then, \eqref{eq:3D212c} becomes
\begin{equation}
\label{eq:3D212d}
 u_k ({\mathbf a}_k) = 0,
\quad k=1,2, \ldots, n\;.
\end{equation} 

The function $u_k$ is written exactly by \eqref{eq:Poincare2}-\eqref{eq:PoiD}. Consider the case of equal radii $r_k=r_0$ and write $u_k(\mathbf x)$ explicitly  up to  $O(r_0^7)$
{\footnotesize 
\begin{align}
\label{eq:PoiDc}
\notag
u_k(\mathbf x) = x_j-c_k
-\sum_{m} \frac{r_0 (a_{mj}-c_m)}{|\mathbf x- \mathbf a_m|}    + \sum_{\substack{m,l}} \frac{r_0^2 (a_{lj}-c_l)}{|\mathbf x- \mathbf a_m||\mathbf a_m- \mathbf a_l|}     
 -\sum_{m} \frac{r_0^3 (x_j-a_{mj})}{|\mathbf x- \mathbf a_m|^3}\\
-\sum_{\substack{m,k,l,s}} \frac{r_0^3 (a_{sj}-c_s)}{|\mathbf x- \mathbf a_m||\mathbf a_m- \mathbf a_l| |\mathbf a_l- \mathbf a_s|}     -\sum_{\substack{m,l}} \frac{r_0^4 (a_{lj}-c_l)  \sum_{i=1}^3 (x_i-a_{mi})(a_{mi}-a_{li}) }{|\mathbf x- \mathbf a_m|^3|\mathbf a_m- \mathbf a_l|^3} \\
\notag
+\sum_{\substack{m,l}} \frac{r_0^4 (a_{mj}-a_{lj})}{|\mathbf x- \mathbf a_m||\mathbf a_m- \mathbf a_l|^3}     +\sum_{\substack{m,l,s,t}} \frac{r_0^4 (a_{tj}-c_t)}{|\mathbf x- \mathbf a_m||\mathbf a_m- \mathbf a_l| |\mathbf a_l- \mathbf a_s| |\mathbf a_s- \mathbf a_t|}\\
\notag
-\sum_{\substack{m,l,s,t,w}} \frac{r_0^5 (a_{wj}-c_{w})}{|\mathbf x- \mathbf a_m||\mathbf a_m- \mathbf a_l| |\mathbf a_l- \mathbf a_s| |\mathbf a_s- \mathbf a_t| |\mathbf a_t- \mathbf a_w|}      -\sum_{\substack{m,l,s}} \frac{r_0^5 (a_{lj}-a_{sj})}{|\mathbf x- \mathbf a_m||\mathbf a_m- \mathbf a_l| |\mathbf a_l- \mathbf a_s|^3} \\
\notag
+\sum_{\substack{m,l,s}} \frac{r_0^5 (a_{sj}-c_s) \sum_{i=1}^3 (x_i-a_{mi})(a_{mi}-a_{li})}{|\mathbf x- \mathbf a_m|^3|\mathbf a_m- \mathbf a_l|^3|\mathbf a_l- \mathbf a_s|}
+\sum_{\substack{m,l,s}} \frac{r_0^5 (a_{sj}-c_s)  \sum_{i=1}^3 (a_{mi}-a_{li})(a_{li}-a_{si})}{|\mathbf x- \mathbf a_m||\mathbf a_m- \mathbf a_l|^3|\mathbf a_l- \mathbf a_s|^3}\\
\notag
-\sum_{\substack{m,l}} \frac{3r_0^6 (a_{lj}-c_l)  \sum_{i=1}^3 (x_i-a_{mi})(a_{mi}-a_{li})}{2|\mathbf x- \mathbf a_m|^5|\mathbf a_m- \mathbf a_l|^5}      +\sum_{\substack{m,l}} \frac{r_0^6 (x_j-a_{mj})}{|\mathbf x- \mathbf a_m|^3|\mathbf a_m- \mathbf a_l|^3} \\
\notag
-\sum_{\substack{m,l}} \frac{r_0^6 (a_{lj}-c_l)  \sum_{i=1}^3 (x_i-a_{mi})^2}{2|\mathbf x- \mathbf a_m|^5|\mathbf a_m- \mathbf a_l|^3}       -\sum_{\substack{m,l}} \frac{3r_0^6 (a_{mj}-a_{lj}) \sum_{i=1}^3 (x_i-a_{mi})(a_{mi}-a_{li})}{|\mathbf x- \mathbf a_m|^3|\mathbf a_m- \mathbf a_l|^5} \\
\notag
-\sum_{\substack{m,l,s,t}} \frac{r_0^6 (a_{tj}-c_t)  \sum_{i=1}^3 (x_i-a_{mi})(a_{mi}-a_{li})}{|\mathbf x- \mathbf a_m|^3|\mathbf a_m- \mathbf a_l|^3|\mathbf a_l- \mathbf a_s| |\mathbf a_s- \mathbf a_t|}     -\sum_{\substack{m,l,s,t}} \frac{r_0^6 (a_{tj}-c_t)  \sum_{i=1}^3 (a_{mi}-a_{li})(a_{li}-a_{si})}{|\mathbf x- \mathbf a_m| |\mathbf a_m- \mathbf a_l|^3|\mathbf a_l- \mathbf a_s|^3 |\mathbf a_s- \mathbf a_t|} \\
\notag
+\sum_{\substack{m,l,s,t}} \frac{r_0^6 (a_{sj}-a_{tj})}{|\mathbf x- \mathbf a_m||\mathbf a_m- \mathbf a_l| |\mathbf a_l- \mathbf a_s| |\mathbf a_s- \mathbf a_t|^3}    -\sum_{\substack{m,l,s,t}} \frac{r_0^6 (a_{tj}-c_t)  \sum_{i=1}^3 (a_{li}-a_{si})(a_{si}-a_{ti})}{|\mathbf x- \mathbf a_m| |\mathbf a_m- \mathbf a_l| |\mathbf a_l- \mathbf a_s|^3 |\mathbf a_s- \mathbf a_t|^3} \\
\notag
+\sum_{\substack{m,l,s,t,w,v}} \frac{r_0^6 (a_{vj}-c_{v})}{|\mathbf x- \mathbf a_m||\mathbf a_m- \mathbf a_l| |\mathbf a_l- \mathbf a_s| |\mathbf a_s- \mathbf a_t| |\mathbf a_t- \mathbf a_w| |\mathbf a_w- \mathbf a_v|} 
+ O(r_0^7)\;.
\end{align}
}

Introduce the designation $z_k:=a_{kj}-c_k$ and substitute $\mathbf x=\mathbf a_k$ into \eqref{eq:PoiDc}. Then, \eqref{eq:3D212d} becomes
\begin{align}
\label{eq:PoiDc3}
z_k
- r_0^3 \sum_{m \neq k} \frac{a_{kj}-a_{mj}}{|\mathbf a_k- \mathbf a_m|^3}
-\sum_{m \neq k} \frac{r_0 z_m}{|\mathbf a_k - \mathbf a_k|} 
+ \sum_{\substack{m \neq k \\ m\neq l}} \frac{r_0^2 z_l}{|\mathbf a_k- \mathbf a_m||\mathbf a_m- \mathbf a_l|}   
\\
\notag
-\sum_{\substack{m \neq k \\ l\neq m\\s\neq l}} \frac{r_0^3 z_s}{|\mathbf a_k- \mathbf a_m||\mathbf a_m- \mathbf a_l| |\mathbf a_l- \mathbf a_s|}  = O(r_0^4), \quad k=1,2,\cdots,n\;.
\end{align}
It can be easily obtained from \eqref{eq:PoiDc3} that
\begin{equation}
\label{eq:3Dc8}
c_k = a_{kj}- r_0^3 \sum_{m \neq k} \frac{a_{kj}-a_{mj}}{|\mathbf a_k- \mathbf a_m|^3}+ O(r_0^4), \quad k=1,2,\cdots,n\;.
\end{equation}
Substitute \eqref{eq:3Dc8}, computed up to $O(r_0^7)$, into \eqref{eq:PoiDc}
\begin{align}
\label{eq:PoiDcf}
\notag
u_k(\mathbf x) = x_j-a_{kj}+ r_0^3 \sum_{m \neq k} \frac{a_{kj}-a_{mj}}{|\mathbf a_k- \mathbf a_m|^3}
-\sum_{m \neq k}\frac{r_0^3 (x_j-a_{mj})}{|\mathbf x- \mathbf a_m|^3} \\
 - \sum_{\substack{m \neq k \\ l\neq m}} \frac{r_0^6(a_{kj}-a_{mj})}{|\mathbf a_k- \mathbf a_m|^3|\mathbf a_m- \mathbf a_l|^3} 
 + \sum_{\substack{m \neq k \\ l\neq m}} \frac{r_0^6(x_j-a_{mj})}{|\mathbf x- \mathbf a_m|^3|\mathbf a_m- \mathbf a_l|^3}   \\
\notag
-\sum_{\substack{m \neq k \\ l\neq m}} \frac{3r_0^6 (a_{mj}-a_{lj}) \sum_{i=1}^3 (x_i-a_{mi})(a_{mi}-a_{li}) }{|\mathbf x- \mathbf a_m|^3|\mathbf a_m- \mathbf a_l|^5}  
\\
\notag
+\sum_{\substack{m \neq k \\ l\neq m}} \frac{3r_0^6 (a_{mj}-a_{lj}) \sum_{i=1}^3 (a_{ki}-a_{mi})(a_{mi}-a_{li}) }{|\mathbf a_k- \mathbf a_m|^3|\mathbf a_m- \mathbf a_l|^5}  
+ O(r_0^7)\;.
\end{align}

This is the main analytical approximate formula which will be used below to compute the effective conductivity tensor.

\section{Averaged conductivity}
\label{subsec22}
The countable set of centers forms the lattice described in Sec.\ref{sec:intr} and can be ordered in the following way $\mathcal A=\{\mathbf a_k+\sum_{s=1,2,3} m_s \boldsymbol{\omega_s}\}$ . Here, $(m_1,m_2,m_3) \in \mathbb Z^3$ is the number of the cell; the points $\mathbf a_k$ ($k=1,2,\ldots,N$) belong to the $\mathbf 0$-cell. 

First, we consider the finite number of inclusions $n=NM$  in the space $\mathbb R^3$ using the linear order numeration $\mathbf a_k$ ($k=1,2,\ldots,n$), i.e., $M$ is the number of cells and the fixed number $N$ is the number of inclusions per cell. We consider equal spheres, i.e., $r_k=r_0$ for simplicity. 
Introduce the value 
\begin{equation}
\label{eq:3Db1}
\lambda_{ij}^{(n)}= \delta_{ij}+  \frac 1M \sum_{k=1}^n   \int_{\partial D_k}(x_i-a_{ki}) \frac{\partial u}{\partial \mathbf n} ds\;. 
\end{equation} 
Then, the component of the effective conductivity tensor $\lambda_{ij}$ can be estimated by formula
\begin{equation}
\label{eq:3Db3}
\lambda_{ij} = \lim_{n\to \infty} \lambda_{ij}^{(n)}+O(\lambda_1^{-1})\;.
\end{equation}
Following the scheme \cite{DM} we use the Eisenstein summation in the limit $n\to \infty$ $\Longleftrightarrow \{M\to \infty, N$ is fixed\}. The subscript $j=1,2,3$ denotes the potential jump per cell along the axis $x_j$ in the triple periodic problem.  

It follows from \eqref{eq:3D2-15} that for every fixed $k$
\begin{equation}
\label{eq:3Dc1}
\int_{\partial D_k}(x_i-a_{k1}) \frac{\partial u}{\partial \mathbf n} ds =
\int_{\partial D_k}\frac{x_i-a_{k1}}{r_0}\; u_k ds+2
\int_{\partial D_k}(x_i-a_{ki}) \frac{\partial u_k}{\partial \mathbf n} ds\;.
\end{equation}
The first integral can be calculated by the Ostrogradsky-Gauss formula 
\begin{equation}
\label{eq:3Dc7}
\int_{\partial D_k}\frac{x_i-a_{ki}}{r_0}\; u_k ds =\int_{D_k}\frac{\partial u_k}{\partial x_i} d\mathbf x\;,
\end{equation}
since the $i$th component of the unit outward normal vector to the sphere $\partial D_k$ is equal to  $\frac{x_i-a_{ki}}{r_k}$. Let $|D_k|$ denote the volume of the ball $D_k$. Application of the mean value theorem for harmonic functions to \eqref{eq:3Dc7} yields
\begin{equation}
\label{eq:3Dc8b}
\int_{D_k} \frac{\partial u_k}{\partial x_i} d \mathbf x = |D_k| \frac{\partial u_k}{\partial x_i} (\mathbf a_k,n)\;.
\end{equation}
 
Green's identity 
and the mean value theorem imply that
\begin{equation}
\label{eq:3Dc9}
\int_{\partial D_k}(x_i-a_{ki}) \frac{\partial u_k}{\partial \mathbf n} ds = |D_k| \; \frac{\partial u_k}{\partial x_i} (\mathbf a_k,n)\;.
\end{equation}
Substitution of \eqref{eq:3Dc7}-\eqref{eq:3Dc9} into \eqref{eq:3Db1}, \eqref{eq:3Dc1} yields
\begin{equation}
\label{eq:3Db2}
\lambda_{ij}^{(n)}=\delta_{ij}+  \frac 3M \sum_{k=1}^n   |D_k|\; \frac{\partial u_k}{\partial x_i} (\mathbf a_k,n)\;,
\end{equation}
where the value $\frac{\partial u_k}{\partial x_1}(\mathbf a_k)$ can be calculated by the approximation \eqref{eq:PoiDcf}. When the external flux is applied in the $x_1$-direction, i.e., $j=1$, we have

{\footnotesize 
\begin{align}
\notag
\frac{\partial u_k}{\partial x_1}(\mathbf a_k,n) =  1 &
+r_0^3 \sum_{m \neq k}\frac{ 2(a_{k1}-a_{m1})^2-(a_{k2}-a_{m2})^2-(a_{k3}-a_{m3})^2}{|\mathbf a_k- \mathbf a_m|^5}
\\
\notag
& +r_0^6 \sum_{m,s}\frac{[2(a_{k_1}-a_{m_1})^2 -(a_{k_2}-a_{m_2})^2-(a_{k_3}-a_{m_3})^2] }{ |\mathbf a_k- \mathbf a_m|^5 |\mathbf a_m- \mathbf a_s|{}^5 } \\
\label{eq:PoiDcg}
& \times [2(a_{m_1}-a_{s_1})^2 -(a_{m_2}-a_{s_2})^2-(a_{m_3}-a_{s_3})^2]
\\ \notag
& +9 r_0^6 \sum_{m,s}\frac{(a_{k_1}-a_{m_1}) (a_{m_1}-a_{s_1}) (a_{k_2}-a_{m_2}) (a_{m_2}-a_{s_2})}{ |\mathbf a_k- \mathbf a_m|^5 |\mathbf a_m- \mathbf a_s|{}^5 }
\\ \notag
& +9 r_0^6 \sum_{m,s}\frac{(a_{k_1}-a_{m_1}) (a_{m_1}-a_{s_1}) (a_{k_3}-a_{m_3}) (a_{m_3}-a_{s_3})}{ |\mathbf a_k- \mathbf a_m|^5 |\mathbf a_m- \mathbf a_s|{}^5 } + O(r_0^7) \;,
\end{align}
}
where the double sum 
$
\sum_{m,s}:=
\sum_{m \neq k} \sum_{s \neq m}\;.
$
{\footnotesize 
\begin{align}
\label{eq:PoiDcg2}
\frac{\partial u_k}{\partial x_2}(\mathbf a_k,n) & =  3r_0^3 \sum_{m,s}\frac{(a_{k_1}-a_{m_1})(a_{k_2}-a_{m_2})}{|\mathbf a_k- \mathbf a_m|^5}\\
\notag
& +3r_0^6 \sum_{m,s}\frac{(a_{k_1}-a_{m_1})(a_{k_2}-a_{m_2})[2(a_{m_1}-a_{s_1})^2 -(a_{m_2}-a_{s_2})^2-(a_{m_3}-a_{s_3})^2]}{|\mathbf a_k- \mathbf a_m|^5 |\mathbf a_m- \mathbf a_s|^5}
\\
\notag
& +3r_0^6 \sum_{m,s}\frac{(a_{m_1}-a_{s_1})(a_{m_2}-a_{s_2})[-(a_{k_1}-a_{m_1})^2 +2(a_{k_2}-a_{m_2})^2-(a_{k_3}-a_{m_3})^2]}{|\mathbf a_k- \mathbf a_m|^5 |\mathbf a_m- \mathbf a_s|^5}
\\
\notag
& +9r_0^6 \sum_{m,s}\frac{(a_{k_2}-a_{m_2})(a_{k_3}-a_{m_3})(a_{m_1}-a_{s_1})(a_{m_3}-a_{s_3})}{|\mathbf a_k- \mathbf a_m|^5 |\mathbf a_m- \mathbf a_s|^5} + O(r_0^7)\;.
\end{align}
}
{\footnotesize 
\begin{align}
\label{eq:PoiDcg3}
\frac{\partial u_k}{\partial x_3}(\mathbf a_k,n) & =  3r_0^3 \sum_{m,s}\frac{(a_{k_1}-a_{m_1})(a_{k_3}-a_{m_3})}{|\mathbf a_k- \mathbf a_m|^5}\\
\notag
& +3r_0^6 \sum_{m,s}\frac{(a_{k_1}-a_{m_1})(a_{k_3}-a_{m_3})}{|\mathbf a_k- \mathbf a_m|^5 |\mathbf a_m- \mathbf a_s|^5}[2(a_{m_1}-a_{s_1})^2 -(a_{m_2}-a_{s_2})^2-(a_{m_3}-a_{s_3})^2]\\
\notag
& +3r_0^6 \sum_{m,s}\frac{(a_{m_1}-a_{s_1})(a_{m_3}-a_{s_3})[-(a_{k_1}-a_{m_1})^2 -(a_{k_2}-a_{m_2})^2+2(a_{k_3}-a_{m_3})^2]}{|\mathbf a_k- \mathbf a_m|^5 |\mathbf a_m- \mathbf a_s|^5}
\\
\notag
& +9r_0^6 \sum_{m,s}\frac{(a_{k_2}-a_{m_2})(a_{k3}-a_{m_3})(a_{m_1}-a_{s_1})(a_{m_2}-a_{s_2})}{|\mathbf a_k- \mathbf a_m|^5 |\mathbf a_m- \mathbf a_s|^5} + O(r_0^7)\;.
\end{align}
}

\section{Effective conductivity}
\label{sub:eff}
Consider a regular infinite array of spheres with $N=1$ and $M=n$, i.e., one sphere per periodicity cell. Rayleigh \cite{Rayleigh} calculated the limit (see formula after (64) in his paper \cite{Rayleigh}) having used the Eisenstein summation 
\begin{equation}
\label{eq:PoiDch}
\lim_{M\to \infty} \frac 1M \sum^M_{\substack{m=1 \\ m\neq k}} \frac{2 (a_{k_1}-a_{m_1})^2-(a_{k_2}-a_{m_2})^2-(a_{k_3}-a_{m_3})^2}{|\mathbf a_k- \mathbf a_m|^5} = \frac 43 \pi\;.
\end{equation}
Substitution of \eqref{eq:PoiDch} into \eqref{eq:PoiDcg} and further into the limit expression \eqref{eq:3Db2} yields the effective conductivity of the simple cubic array of sphere
\begin{equation}
\label{eq:3Db20}
\lambda_{11} = 1+ 3f+3f^2+O(r_0^7)\;,
\end{equation}
where $f= \frac 43 \pi r_0^3$ denotes the concentration of spheres (one sphere per unit cubic cell). 

For general $N$ the formula \eqref{eq:3Db20} is replaced by the following relation
\begin{equation}
\label{eq:3Db21}
\lambda_{11} = 1+ 3f+3 f^2 \;\frac 3{4\pi}e_{11}+O(r_0^7)\;,
\end{equation}
where 
\begin{equation}
\label{eq:3Db22}
e_{11} = \frac{1}{N^2} \sum_{k,m = 1}^N E_{11}(\mathbf a_k -\mathbf a_m)\;,
\end{equation}
\begin{equation}
\label{eq:3Db23}
E_{11}(\mathbf x) = \sum_{\mathbf R} \frac{ 2(x_1-R_1)^2-(x_2-R_2)^2-(x_3-R_3)^2}{|\mathbf x-\mathbf R|^5}\;. 
\end{equation}
The vectors $\mathbf R=(R_1,R_2,R_3) = \sum_{i=1,2,3} m_i \boldsymbol{\omega}_i$ $( m_i \in \mathbb Z)$ generate the simple cubic lattice. The Eisenstein summation is used in \eqref{eq:3Db23} and it is assumed for shortness that $E_{11}(\mathbf 0):=\frac 43 \pi$ when $\mathbf a_k=\mathbf a_m$ is taken in \eqref{eq:3Db22}.

The following relation  holds $E_{11} = -\frac{\partial E_1}{\partial x_1}$, where $E_1(\mathbf x)$ is the first coordinate of the function (8.4.77) introduced in \cite{GMN}
\begin{equation}
\label{eq:3De1}
E_{1}(\mathbf x) = \sum_{\mathbf R} \frac{x_1-R_1}{|\mathbf x-\mathbf R|^3}\;. 
\end{equation}
This function is related to Berdichevsky's function $\wp_{11}(\mathbf x)$ considered in \cite{Berdichevsky}\footnote{The 3D $\wp$-functions are discussed only in the Russian edition \cite{Berdichevsky}.} (see also \cite[Sec.3, Chapter 8]{GMN})
\begin{equation}
\label{eq:3De2}
E_{11}(\mathbf x) = \frac 43 \pi +\wp_{11}(\mathbf x)\;. 
\end{equation}
The function $\wp_{11}(\mathbf x)$ was introduced in \cite{Berdichevsky} through the absolutely convergent series
\begin{align}
\label{eq:3DB11}
& \wp_{11}(\mathbf x) = \frac{ 2x_1^2-x_2^2-x_3^2}{|\mathbf x|^5}
\\ \notag
& +\sum\nolimits'_{\mathbf R} \left[-\frac{1}{|\mathbf x-\mathbf R|^3}+\frac{1}{|\mathbf R|^3}+3 \left(\frac{ (x_1-R_1)^2}{|\mathbf x-\mathbf R|^5}- \frac{ R_1^2}{|\mathbf R|^5}\right) \right]\;, 
\end{align}
where $\mathbf R$ runs over $\mathbb Z^3$ in the infinite sum $\sum\nolimits'_{\mathbf R}$ except $\mathbf R = \mathbf 0$. 
A power series of $E_{11}(\mathbf x)$ on the coordinates of $\mathbf x$ can be applied to effectively compute the values $E_{11}(\mathbf a_k -\mathbf a_m)$. Using the standard multidimensional Taylor expansion in the coordinates of $\mathbf x$ we obtain
\begin{align}
\label{eq:3De3}
E_{11}(\mathbf x) = \frac 43 \pi +  \frac{ 2x_1^2-x_2^2-x_3^2}{|\mathbf x|^5}
 +
\sum\nolimits'_{\mathbf R} \frac{3 R_1 \left(2 R_1^2 - 3 \left(R_2^2+R_3^2\right)\right)}{|\mathbf R|^7} \;x_1+ 
\\
\notag
\sum\nolimits'_{\mathbf R} \frac{3 R_2 \left(4   R_1^2 -R_2^2 -R_3^2\right)}{|\mathbf R|^7}\;x_2 + \sum\nolimits'_{\mathbf R} \frac{3 R_3 \left(4R_1^2 -R_2^2 -R_3^2\right)}{|\mathbf R|^7}\;x_3+\ldots\;, 
\end{align}
where, for instance,
\begin{equation}
\label{eq:3De33}
\sum\nolimits'_{\mathbf R} \left[\frac{1}{|\mathbf R|^3}-\frac{1}{|\mathbf x-\mathbf R|^3}\right]= 3\sum\nolimits'_{\mathbf R} \frac{R_1 x_1+R_2 x_2+R_3 x_3}{|\mathbf R|^7}+ \ldots \;.
\end{equation}
All the sums $\sum\nolimits'_{\mathbf R}$ in \eqref{eq:3De3} are at least of order $|\mathbf R|^{-4}$, hence, they converge absolutely \cite{Berdichevsky}. In this case we can change the order of summation and use their symmetry in $R_i$. In particular, this implies that the coefficients in $x_j$ vanish.

The function $E_{12} = -\frac{\partial E_1}{\partial x_2}$ coincides with Berdichevsky's function $\wp_{12}(\mathbf x)$ \cite{Berdichevsky} and can be presented as the Eisenstein series 
\begin{equation}
\label{eq:3De12}
E_{12}(\mathbf x) = 3\sum_{\mathbf R} \frac{ (x_1-R_1) (x_2-R_2)}{|\mathbf x-\mathbf R|^5}\;. 
\end{equation}
The function $\wp_{12}(\mathbf x)$ was expanded in \cite{Berdichevsky} into the absolutely convergent series
\begin{equation}
\label{eq:3DB12}
\wp_{12}(\mathbf x) =3\left\{ \frac{x_1 x_2}{|\mathbf x|^5}+\sum\nolimits'_{\mathbf R} \left[ \frac{ (x_1-R_1)(x_2-R_2)}{|\mathbf x-\mathbf R|^5}-\frac{R_1R_2)}{|\mathbf R|^5} \right] \right\}\;. 
\end{equation}
The functions $E_{21}(\mathbf x)$ and $\wp_{21}(\mathbf x)$ are introduced analogously to \eqref{eq:3De12} and \eqref{eq:3DB12} by the interchange of subscripts $1$ and $2$. It follows from the absolute convergence of \eqref{eq:3DB12} that
\begin{equation}
\label{eq:3D1221}
\wp_{12}(\mathbf x) =\wp_{21}(\mathbf x), \; \mbox{hence}, \;
E_{12}(\mathbf x) = E_{21}(\mathbf x)\;.
\end{equation}

The function $E_{13} = -\frac{\partial E_1}{\partial x_3}$ is similar to $E_{12}(\mathbf x)$ and can be calculated analogously by replacement the subscript $2$ by $3$. The functions $E_{1j}$ determine the structural sums similar to \eqref{eq:3Db22} 
\begin{equation} 
\label{eq:3Db22a}
e_{1j} = \frac{1}{N^2} \sum_{k,m = 1}^N E_{1j}(\mathbf a_k -\mathbf a_m)\;, \quad (j=2,3)\;.
\end{equation}

Introduce the absolutely convergent sums ($i_1+i_2+\ldots+i_m \leq \ell-4$) symmetric with respect to $R_i$
\begin{equation}
\label{eq:3De5}
e_{\ell}^{(i_1,i_2,i_3)} = \sum\nolimits'_{\mathbf R} \frac{R_1^{i_1}R_2^{i_1}R_3^{i_3}}{|\mathbf R|^\ell}\;. 
\end{equation}
Note that $e_{\ell}^{(i_1,i_2,i_3)} = 0$ when at least one superscript $i_k$ is odd. Then, \eqref{eq:3De3} is reduced to
\begin{align}
\label{eq:3De4}
E_{11}(\mathbf x) = \frac 43 \pi + \frac{ 2x_1^2-x_2^2-x_3^2}{|\mathbf x|^5} -\frac{21}{2} \left(2 x_1^2-x_2^2-x_3^2\right) \left(3 e_9^{2,2,0}-e_9^{4,0,0}\right)\\ \notag
 +\frac{45}{8} \left(2 x_1^4-6
   \left(x_2^2+x_3^2\right) x_1^2-x_2^4-x_3^4 +12
   x_2^2 x_3^2\right) \\ \notag  \times  \left(30
   e_{13}^{2,2,2}-15
   e_{13}^{4,2,0}+e_{13}^{6,0,0}\right) \\ \notag
 +\frac{693}{16} \left(2 x_1^6-15
   \left(x_2^2+x_3^2\right) x_1^4 +15
   \left(x_2^4+x_3^4\right)
   x_1^2 -x_2^6-x_3^6\right) \\ \notag
   \times \left(35
   e_{17}^{4,4,0}-28
   e_{17}^{6,2,0}+e_{17}^{8,0,0}\right) \\ \notag
 -\frac{585}{128} \left(10 x_1^8-140
   \left(x_2^2+x_3^2\right) x_1^6+70
   \left(x_2^4+24 x_3^2 x_2^2+x_3^4\right)
   x_1^4 \right. \\ \notag
   \left. +28 \left(x_2^6-30 x_3^2 x_2^4-30 x_3^4
   x_2^2+x_3^6\right) x_1^2-5 x_2^8-5 x_3^8+112
   x_2^2 x_3^6 \right. \\ \notag
   \left. -140 x_2^4 x_3^4+112 x_2^6
   x_3^2\right) \left(630 e_{21}^{4,4,2}-504
   e_{21}^{6,2,2} \right. \\ \notag
   \left.-42 e_{21}^{6,4,0}+45
   e_{21}^{8,2,0}-e_{21}^{10,0,0}\right)+\ldots\;.
\end{align}
 
The function $E_{12}(\mathbf x)$ is calculated by the similar formula
\begin{align}
\label{eq:3De412}
E_{12}(\mathbf x) =  \frac{3 x_1 x_2}{|\mathbf x|^5} + 21 x_1 x_2 \left(3
   e_9^{2,2,0}-e_9^{4,0,0}\right) \\ \notag
  -\frac{45}{2} x_1 x_2 \left(x_1^2+x_2^2-6
   x_3^2\right) \left(30 e_{13}^{2,2,2}-15
   e_{13}^{4,2,0}+e_{13}^{6,0,0}\right) \\ \notag
 -\frac{693}{8} x_1 x_2 \left(3 x_1^4-10 x_2^2
   x_1^2+3 x_2^4\right) \left(35
   e_{17}^{4,4,0}-28
   e_{17}^{6,2,0}+e_{17}^{8,0,0}\right) \\ \notag
 +\frac{585}{16} x_1 x_2 \left(5 x_1^6-7
   \left(x_2^2+12 x_3^2\right) x_1^4-7
   \left(x_2^4-20 x_3^2 x_2^2-10 x_3^4\right)
   x_1^2 \right. \\ \notag
  \left. +5 x_2^6-28 x_3^6+70 x_2^2 x_3^4-84 x_2^4
   x_3^2\right) \left(630 e_{21}^{4,4,2}-504
   e_{21}^{6,2,2} \right. \\ \notag
  \left. -42 e_{21}^{6,4,0}+45
   e_{21}^{8,2,0}-e_{21}^{10,0,0}\right) +\ldots\;.
\end{align}
It is convenient for computations to replace combinations of $e_{\ell}^{(i_1,i_2,i_3)}$ with Coulombic lattice sums \cite{Huang1999} (see Appendix~\ref{sec:lattSums})
\begin{align}
 \label{eq:E11L}
E_{11}(\mathbf x) = \frac 43 \pi + \frac{ 2x_1^2-x_2^2-x_3^2}{|\mathbf x|^5} + 6{\cal L}_4^0\left(2 x_1^2- \left(x_2^2+x_3^2\right)\right) \\ \notag
   + 15 {\cal L}_6^0 \left(2 x_1^4-6
   \left(x_2^2+x_3^2\right) x_1^2-x_2^4-x_3^4+12
   x_2^2 x_3^2\right) \\ \notag
  + 28 {\cal L}_8^0 \left(2 x_1^6-15 \left(x_2^2+x_3^2\right) x_1^4+15
   \left(x_2^4+x_3^4\right)
   x_1^2-x_2^6-x_3^6\right)  \\ \notag
    + 9 {\cal L}_{10}^0
   \left(10 x_1^8-140 \left(x_2^2+x_3^2\right)
   x_1^6+70 \left(x_2^4+24 x_3^2
   x_2^2+x_3^4\right) x_1^4 \right. \\ \notag
  \left. +28 \left(x_2^6-30
   x_3^2 x_2^4-30 x_3^4 x_2^2+x_3^6\right) x_1^2 -5(
   x_2^8- x_3^8) \right. \\ \notag
  \left. +112 (x_2^2 x_3^6 + x_2^6 x_3^2) -140 x_2^4
   x_3^4\right)  + \ldots \;,
\end{align}
\begin{align}
 \label{eq:E12L}
E_{12}(\mathbf x) = \frac{3 x_1 x_2}{|\mathbf x|^5} -12 {\cal L}_4^0 x_1 x_2  -60 {\cal L}_6^0 x_1 \left(x_1^2+x_2^2-6
   x_3^2\right) x_2   \\ \notag
   -56 {\cal L}_8^0 x_1 \left(3 x_1^4-10
   x_2^2 x_1^2+3 x_2^4\right) x_2
    -72  {\cal L}_{10}^0 x_1 \left(5
   x_1^6-7 \left(x_2^2+12 x_3^2\right) x_1^4 \right. \\ \notag
   \left. -7
   \left(x_2^4-20 x_3^2 x_2^2-10 x_3^4\right)
   x_1^2+5 x_2^6-28 x_3^6+70 x_2^2 x_3^4-84 x_2^4
   x_3^2\right) x_2 + \ldots\;.
\end{align}
The function $E_{13}(\mathbf x)$ can be obtained from \eqref{eq:E12L} by the interchange $x_2$ and $x_3$.

Introduce the discrete spatial convolution sums constructed by the sums \eqref{eq:PoiDcg}-\eqref{eq:PoiDcg3}
\begin{equation}
\label{eq:3De1111}
e_{ij*pl}=\frac{1}{N^3} \sum_{k,m,s=1}^N E_{ij}(\mathbf a_k- \mathbf a_m)E_{pl}(\mathbf a_m- \mathbf a_s)\;.
\end{equation}
The values $e_{ij}$ and $e_{ij*pl}$ will called the {\it structural sums}. They generalize the classic lattice sums constructed for regular arrays. The approximation \eqref{eq:3Db21} is extended by application of \eqref{eq:PoiDcg} 
\begin{equation}
\label{eq:3Def6}
\lambda_{11} = 1+ 3f+3 f^2 \;\frac 3{4\pi}e_{11}+3 f^3 \left(\frac 3{4\pi}\right)^2
[e_{11*11}+3(e_{12*12}+e_{13*13})] +O(f^{\frac{10}3})\;.
\end{equation}

It follows from \eqref{eq:PoiDcg2} that
\begin{equation}
\label{eq:3Def6i}
\lambda_{12} = 9f^2 \left[\frac 3{4\pi}e_{12}+ f\;\left(\frac 3{4\pi}\right)^2 (e_{12*11}+ e_{12*22}+e_{13*13}+e_{23*13})\right] +O(f^{\frac{10}3})\;.
\end{equation}
The component $\lambda_{13}$ can be derived from \eqref{eq:PoiDcg3}. Ultimately, it is written by the interchange of subscripts $2$ and $3$ in \eqref{eq:3Def6i}.

\begin{remark}
\label{rem:1}
In the case of the simple cubic array when $N=1$, the convolution \eqref{eq:3De1111} is simplified to $e_{11*11}=\left(\frac{4\pi}3\right)^2$ and $e_{12*12}=e_{13*13}=0$.
The effective conductivity of the simple cubic array can be calculated by the Clausius-Mossotti approximation \eqref{eq:CMA} for $\beta=1$ 
\begin{equation}
\label{eq:3DCMA}
\lambda_{11} = \frac{1+ 2f}{1-f}+O(f^{4})
\end{equation}
what corresponds to the general formula \eqref{eq:3Def6}. The most advanced formula for the simple cubic array was established by Berdichevsky \cite[formula (11.80)]{Berdichevsky} 
\begin{equation}
\label{eq:3DCMAB}
\lambda_{e} = \frac{1+ 2f}{1-f}+ \frac{3.913f^{\frac{13}3}}{(1-f)^2}+ \frac{1.469f^{\frac{17}3}}{(1-f)^2} +O(f^{6})\;,
\end{equation}
where $\lambda_{e}=\lambda_{11}$ is the scalar effective conductivity of the simple cubic array.
\end{remark}

\section{Numerical results for random composites}
\label{sect:numRes}
We now proceed to apply \eqref{eq:3De1111} to numerical estimation of the effective conductivity of random macroscopically isotropic composites.
In samples generation we applied Random Sequential Adsorption (RSA) protocol, where consecutive objects are placed randomly in the cell, rejecting those that overlap previously absorbed ones. We generated 10 samples of the RSA distribution with $N=1000$ balls each. The concentration $f=0.3$ is fixed during the generation. Other concentrations about $f=0.3$ give practically the same result. Such a sample is displayed in Fig.\ref{fig:sample}. 

In calculations of~\eqref{eq:3Def6} we used the following approximated values of convergent lattice sums
\begin{equation}
\label{eq:N1}
{\cal L}_4^0 = 3.10822,\; {\cal L}_6^0=0.573329,\; {\cal L}_8^0=3.25929,\; {\cal L}_{10}^0=1.00922\:, 
\end{equation}
computed as partial sums of~\eqref{eq:3De5} for $-250\leq R_k \leq 250$ ($k=1,2,3$) (see Appendix~\ref{sec:lattSums}). Obtained approximations agree to at least five significant digits with known numerical values~\cite{Huang1999}. Numerical results are presented in Table~\ref{tab:numRes}.
\begin{table}[!h]
\center
$
\begin{array}{ccccc}
 e_{11} & e_{11*11} & e_{12*12} & e_{13*13} \\\hline
 4.13295 & 19.5071 & 1.41568 & 1.41133  \\
 4.19783 & 19.6056 & 1.39797 & 1.55950  \\
 4.17890 & 19.2992 & 1.50897 & 1.49774  \\
 4.22268 & 19.5005 & 1.40287 & 1.38913  \\
 4.19561 & 19.6061 & 1.41058 & 1.42387  \\
 4.16198 & 19.0934 & 1.41535 & 1.52156  \\
 4.21459 & 19.6623 & 1.47737 & 1.42095  \\
 4.18675 & 19.5398 & 1.45704 & 1.44189  \\
 4.22531 & 19.5500 & 1.40073 & 1.29346  \\
 4.19565 & 19.3031 & 1.39020 & 1.58077  \\
 & & & & \\
  \overline{e_{11}} &  \overline{e_{11*11}} & \overline{e_{12*12}} & \overline{e_{13*13} }  \\\hline
  4.19122 & 19.4667 & 1.42768 & 1.45402
\end{array}
$
\label{tab:numRes}
\caption{Numerical results for the structural sums (top) and the corresponding mean values (bottom). The bar stands for the mean values.}
\end{table}

The structural sums \eqref{eq:3Db22} and \eqref{eq:3De1111} are special cases of {\it discrete multidimensional convolution of functions} defined in \cite{Naw2018}. In our computations we applied efficient algorithms developed therein. The total time of calculations of results from Table~\ref{tab:numRes}, ran on a standard notebook equipped with Intel Core i7 Processor of 4th generation, was about 30 minutes. 

One can see that $\overline{e_{11}}$ is close to $\frac{4\pi}{3}\approx 4.18879 $. We suggest that  $e_{11} = \frac{4\pi}{3}$ for ideally macroscopically isotropic composites. This hypothesis can be confirmed by the following observations. 
The expansion \eqref{eq:E11L} of $E_{11}(\mathbf x)$ is a linear combination of the even homogeneous polynomials $P_l(x_1,x_2,x_3)$. One can directly check for few $l$ that   
\begin{equation}
\label{eq:3Db40}
P_l(x_1,x_2,x_3)+P_l(x_2,x_1,x_3)+P_l(x_3,x_2,x_1) =0\;.
\end{equation}
We suggest that \eqref{eq:3Db40} holds for all $P_l(x_1,x_2,x_3)$ in the expansion \eqref{eq:E11L}. Consider 24 points 
$$(\pm a_1,\pm a_2,\pm a_3),\;(\pm a_2,\pm a_1,\pm a_3),\;(\pm a_3,\pm a_2,\pm a_1)$$ 
in the $\mathbf 0$-cell generating a macroscopically isotropic structure. If \eqref{eq:3Db40} is true, $e_{11} = \frac{4\pi}{3}$ for these 24 points. Therefore, \eqref{eq:3Def6} for the considered macroscopically isotropic composites yields the scalar effective conductivity    
\begin{equation}
\label{eq:3Def6is}
\lambda_{e} = 1+ 3f+3 f^2+4.80654 f^3+O(f^{\frac{10}3})\;.
\end{equation}

Remark \ref{rem:1} concerning the simple cubic array can be used to improve the polynomial formulae \eqref{eq:3Def6}-\eqref{eq:3Def6i}. These formulae are derived by the method of successive approximations applied to the system of functional equations \eqref{eq:3D37} for $r_m=r_0$. We write exactly all the terms up to $O(r_0^{7})$ taking into account interactions among all the spheres. We shall increase this precision in a future work. But now we can exactly obtain some high order terms in the considered case of macroscopically isotropic composites. If we stay in the $p$th iteration applied to \eqref{eq:3D37} only the term with $r_0^{p}$ we obtain a sequence of terms which leads to the Clausius-Mossotti approximation $1+3f+3f^2+3f^3+\ldots = \frac{1+ 2f}{1-f}$. Berdichevsky's formula \eqref{eq:3DCMAB} takes into account interactions between spheres located in different periodicity cells. The same terms have to be in the formula for random case as in \eqref{eq:3DCMAB}. The above argumentation makes possible to rewrite \eqref{eq:3Def6is} in the asymptotically equivalent form
\begin{equation}
\label{eq:3DCMABi}
\lambda_{e} = \frac{1+ 2f}{1-f}+1.80654 f^3+ \frac{3.913f^{\frac{13}3}}{(1-f)^2}+ \frac{1.469f^{\frac{17}3}}{(1-f)^2} +O(f^{\frac{10}3})\;.
\end{equation}

\section{Application to stir casting process}
\label{sec:Appl}
Properties of dispersed composites are strongly influenced by distributions of inclusions and interactions between them. It is observed in the experimental works \cite{RK, RK2017, Karkri} that composites with the same constitutes, shapes of inclusions and concentrations may have different macroscopic properties. The key of this diversity lies in different distributions of inclusions. Formulae \eqref{eq:3Def6}-\eqref{eq:3Def6i} give us such a possibility to distinguish the geometric structures. This implies wide applications of formulae \eqref{eq:3Def6}-\eqref{eq:3Def6i} to optimal design problems in fabrication of composites, to macroscopic description of polymers and to other problems where the geometry plays the crucial role.   

In the present section, we develop a general effective scheme to distinguish macroscopically isotropic and anisotropic composites obtained in stir casting process. Stir casting is a technological process for the fabrication of Al-SiC and other composites \cite{RK, RK2017}. Below, we develop a method to optimize technological steps involved in 3D stir casting process. It is an extension of the 2D method proposed in \cite{Mit2001}. 

It is established in Sec.\ref{sub:eff} that the normalized effective conductivity tensor for a general distribution of inclusions can be presented in the form
\begin{equation}
\label{eq:scp1}
\Lambda_e =(1+3f)I +f^2 \Lambda^{(2)}+O(f^3), 
\end{equation}
where $I$ denotes the unit tensor. The symmetric tensor $\Lambda^{(2)}$ is explicitly written in terms of the structural sums 
\begin{equation}
\label{eq:scp2}
\Lambda^{(2)} = \frac{9}{4\pi} \left(
\begin{array}{llll}
e_{11} \quad 3e_{12} \quad 3e_{13}
\\
3e_{21} \quad e_{11}^{*} \quad 3e_{23}
\\
3e_{31} \quad 3e_{32} \quad e_{11}^{**}
\end{array}
\right)\;,
\end{equation}
where $e_{ij}$ are given by \eqref{eq:3Db22} and \eqref{eq:3Db22a}. Construction of the components $e_{11}^{*}$ and $e_{11}^{**}$ requires an explanation. The structural sum $e_{11}$ is calculated by equations \eqref{eq:3Db22}-\eqref{eq:3Db23} where the triple periodic function $E_{11}(\mathbf x)$ is defined by the Eisenstein summation. The considered Eisenstein series is defined as a triple symmetric iterated series first summed along the $x_1$--axis and after along the $x_2$-- and $x_3$-- axes. According to the Rayleigh-McPhedran formalism the prime summation along the $x_1$--axis is justified by the applied external flux $(1,0,0)$. This method yields the component $e_{11}$.

The component $e_{11}^{*}$ in the tensor \eqref{eq:scp2} corresponds to the conductivity in the $x_2$--direction when the external flux is applied along the $x_2$--axis. Hence, $e_{11}^{*}$ can be calculated analogously to $e_{11}$ by interchanging the $x_1$-- and $x_2$-- axes. In order to avoid ambiguity in the definition of Eisenstein summation we may introduce $e_{11}^{*}$ using the standard function $E_{11}(\mathbf x)$ defined by \eqref{eq:3Db23}
\begin{equation}
\label{eq:3Db22s}
e_{11}^{*} = \frac{1}{N^2} \sum_{k,m = 1}^N E_{11}(\mathbf a_k^{*} -\mathbf a_m^{*})\;,
\end{equation} 
where $\mathbf a_k^{*}$ is obtained from $\mathbf a_k$ by interchanging the first and second coordinates, i.e. $\mathbf a_k^{*} = (a_{k2},a_{k1},a_{k3})$. The value $e_{11}^{**}$ is calculated analogously through the same Eisenstein series $E_{11}(\mathbf x)$ by interchanging the first and third coordinates.  

We now introduce a 3D anisotropy coefficient following the 2D method \cite{Mit2001}. The deviatoric component of tensor $\Lambda^{(2)}$  has the form 
\begin{equation}
\label{eq:scp4}
\mbox{Dev} \; \Lambda^{(2)} = \frac{3}{4\pi} \left(
\begin{array}{llll}
g_{11} \quad 9e_{12} \quad 9e_{13}
\\
9e_{21} \quad g_{22} \quad 9e_{23}
\\
9e_{31} \quad 9e_{32} \quad g_{33}
\end{array}
\right)\;,
\end{equation}
where 
\begin{equation}
\label{eq:scp5}
g_{11}=2e_{11}-e_{11}^{*}-e_{11}^{**},\;g_{22}=2e_{11}^{*}-e_{11}-e_{11}^{**},\;g_{33}= 2e_{11}^{**}-e_{11}-e_{11}^{*}\;.
\end{equation}  
We introduce the anisotropy scalar coefficient as the absolute value of the determinant of $\Lambda^{(2)}$,
\begin{equation}
\label{eq:scp6}
\kappa = |\det \mbox{Dev} \; \Lambda^{(2)} |\;.
\end{equation} 
One can see that $\kappa \geq 0$ and $\kappa$ vanishes for macroscopically isotropic media.

The anisotropy coefficient \eqref{eq:scp6} yields a powerful method to optimize parameters of stir casting processes for moderate concentrations. It is based on the following steps described in \cite{RK, RK2017} for 2D composites. First, a 3D sample has to be processed to determine the coordinates $\{\mathbf a_1, \mathbf a_2, \ldots \mathbf a_N\}$. Next, the coefficient $\kappa$ is calculated by  \eqref{eq:scp4}-\eqref{eq:scp6} where $e_{ij}$ are exactly given by \eqref{eq:3Db22} and \eqref{eq:3Db22a}. If $\kappa$ is closed to zero, the considered composite is macroscopically isotropic. In the opposite case, the stir casting process should be continued to get a more uniform distribution of inclusions. 

\section{Discussion and conclusion}
\label{sec:concl}
Derivation of analytical (approximate) formulae for the effective properties of composites requires a subtle mathematical study of the conditionally convergent sums arisen in the course of spatial averaging. The first order approximations \eqref{eq:CMA} and \eqref{eq:Einstein} in $f$ were obtained as solutions to the single-inclusion problem. 

For many years it was thought that Maxwell's and Clausius-Mossotti approximations can be systematically and rigorously extended to higher orders in $f$ by taking into account interactions between pairs of spheres, triplets of spheres, and so on. However, it was recently demonstrated \cite{Mit2018, GMN} that the field around a finite cluster of inclusions can yield a correct formula for the effective conductivity only for non-interacting clusters. The higher order term can be properly found only after a subtle study of the conditionally convergent series. In the present paper, the Eisenstein summation is used following \cite{Rayleigh} for a regular cubic lattice. Justification of the Eisenstein summation was given in \cite[Sec.2.4]{McPhedran1} by studying the shape-dependent sums.    
The present paper contains an algorithm to compute the effective conductivity tensor with an arbitrary precision in $f$. The symbolic computations are performed up to  $O(f^{\frac{10}3})$. As a result, the analytical approximate formulae \eqref{eq:3Def6}-\eqref{eq:3Def6i}  are derived for an arbitrary locations of non-overlapping spheres, and \eqref{eq:3DCMABi} for a class of macroscopically isotropic composites.
These formulae explicitly demonstrate the dependence of the effective conductivity tensor on the deterministic and probabilistic distributions of inclusions.

It is interesting to compare our analytical formulae \eqref{eq:3Def6}-\eqref{eq:3Def6i} with others. We leave aside some numerical results for fixed locations of inclusions \cite{Zhang}, since they do not contain the symbolic dependencies on the locations. Allthough such a dependence can be investigated by Monte Carlo methods, it requires a tedious, time-consuming computer simulations. It is rather impossible at the present time to get an accurate results. It is worth noting that our numerical results presented in Sec.\ref{sect:numRes} are obtained for $10$ generated locations with $N=1000$ balls each, by using a standard laptop computer. 

Some authors equate an approximate analytical formula with a {\it model}. Such an approach is misleading, since a mathematical modeling involves the fixed governed equations, interface and boundary conditions. 
Different approxmate formulae/solutions for the mathematical model 
hold under restrictions usually not discussed by authors. 
A serious methodological mistake may follow when intermediate manipulations are valid only within the precision $O(f)$, while the final formula is claimed  
to work with a higher precision, see an explicit example in \cite{Mit2018a}. In particular, it follows from our formulae \eqref{eq:3Def6}-\eqref{eq:3Def6i} that it is impossible to write a universal higher order formula independent on locations of inclusions. Such a universal formula holds only for a limited class of composites with 
non-interacting inclusions, e.g., for a dilute composites and the Hashin-Shtrikman coated sphere assemblage \cite{Cherkaev}.  

It is exactly the case of misundersood precision which is confronted here. All the formulae \eqref{eq:CMA} and \eqref{eq:Jeff} for $\beta=1$, and \eqref{eq:3DCMABi} are the same up to $O(f^2)$. But not always formulae \eqref{eq:CMA} and \eqref{eq:Jeff} can be extended to high concentrations. Moreover, there is a plenty of random distributions. Formula \eqref{eq:3DCMABi} holds up to $O(f^{\frac{10}3})$ for randomly distributed non-overlapping spheres realized with the RSA protocol. Another random distribution yields another values of the structural sums selected in Table \ref{tab:numRes} for the RSA protocol.

Let us address the Hashin-Shtrikman bounds \cite{Milton}. The dependence of the effective conductivity of a random composite in $f$ corresponds to some monotonous curve drawn between the bounds. Such a curve can be  sketched arbitrarily, and it will correspond to some unspecified distribution of inclusions. In the present paper, we deal with a uniform distribution corresponding to a stir-casting process described by the RSA model \cite{RK, RK2017}. The main theoretical requirement to the geometric model consists not only in writing a formula, but in a precise description of the geometrical conditions imposed on deterministic or random locations of inclusions. The rigorous statement and study of this theoretical problem is necessary for the proper approach to various applied problems, e.g., stir-casting process \cite{RK} discussed in Sec.\ref{sec:Appl}. 

The comparison of \eqref{eq:3Def6}-\eqref{eq:3Def6i} and \eqref{eq:3DCMABi} with numerical simulations and experimental results \cite{Karkri, Zhang} demonstrates their agreement for concentrations not exceeding $0.25$. For higher concentrations the results can be different. It is not surprising, since the formulae \eqref{eq:3Def6}-\eqref{eq:3Def6i} explicitly demonstrate the dependence of the effective conductivity on the location of inclusions. The available experimental studies do not include the locations of inclusions, making comparison difficult.

As it is noted above, the formula \eqref{eq:3DCMABi} is derived  
for the mathematical expectation of the effective conductivity over the independent and identically distributed (i.i.d.) non-overlapping inclusions. Even the formula \eqref{eq:3DCMABi} for uniformly distributed non-overlapping balls is not universal, because the effective conductivity depends on the protocol of computer simulations or experimentral stirring, meaning that the very notion of randomness is non-universal \cite{Torquato, GMN, RK, Ryl2014}.  

To summarize, the correct form of the $f^2$ term in \eqref{eq:CMA} is determined. It is rigorously justified for a macroscopically isotropic media with highly conducting spherical inclusions. We explicitly demonstrate that the next $f^3$ term, depends on the deterministic and random locations of inclusions.        

\appendix

\section{Lattice sums}
\label{sec:lattSums}

In three dimensions, Coulombic lattice sums are defined through the following formula~\cite{Huang1999}
\begin{equation}
\label{eq:N2} 
{\cal L}_n^m=\sum\nolimits'_{\mathbf R} A_n^m \left(\frac{\partial}{\partial x_1}+i\frac{\partial}{\partial x_2}\right)^m \left(\frac{\partial}{\partial x_3}\right)^{n-m} \left(\frac{1}{|\mathbf R|}\right)\;, 
\end{equation}
where
\begin{equation}
\label{eq:N3} 
A_n^m=\frac{(-1)^n}{\sqrt{(n-m)!(n+m)!}}\;.
\end{equation}
As an example let us expand the sum ${\cal L}_4^0 $
\begin{equation}
\label{eq:N4}  
{\cal L}_4^0 =  \frac 18 \sum\nolimits'_{\mathbf R}  \frac{3 R_1^4+6 R_2^2 R_1^2-24 R_3^2 R_1^2+3 R_2^4+8 R_3^4-24 R_2^2 R_3^2}{|\mathbf R|^9}\;. 
\end{equation}
Hence, by \eqref{eq:3De5} we have
\begin{equation}
\label{eq:N5} 
{\cal L}_4^0 =  \frac 18 \left( 3e_9^{(4,0,0)}+3e_9^{(0,4,0)}+8e_9^{(0,0,4)}+6e_9^{(2,2,0)}-24e_9^{(2,0,2)}-24e_9^{(0,2,2)} \right)\;. 
\end{equation}
Since the sums \eqref{eq:3De5} are absolutely convergent, the  permutation of indexes $i_k$ do not change their values. Therefore, the considered lattice sum simplifies to
\begin{equation}
\label{eq:N6} 
{\cal L}_4^0 = -\frac{7}{4} \left(3
   e_9^{2,2,0}-e_9^{4,0,0}\right)\;. 
   \end{equation}
Other lattice sums can be simplified by the same method
\begin{align}
{\cal L}_6^0 & =  \frac{3}{8} \left(30 e_{13}^{2,2,2}-15
   e_{13}^{4,2,0}+e_{13}^{6,0,0}\right)\;,\\
   \notag  
{\cal L}_8^0 & =  \frac{99}{64} \left(35 e_{17}^{4,4,0}-28
   e_{17}^{6,2,0}+e_{17}^{8,0,0}\right)\;,\\
   \notag  
{\cal L}_{10}^0 & =     -\frac{65}{128} \left(630 e_{21}^{4,4,2}-504
   e_{21}^{6,2,2}-42 e_{21}^{6,4,0}+45
   e_{21}^{8,2,0}-e_{21}^{10,0,0}\right)\;.
\end{align}

\section{Model of symbolic computations}
\label{sec:symComp}

Let us present a general model of symbolic computations of the solution of the system of functional equations \eqref{eq:3D37} in form of the analytic approximation \eqref{eq:PoiDcf}.
The main difficulty in programming such a procedure is tackling {\it nested} summations and the fact that they are {\it indefinite}, i.e. we do not know in advance the number of functions $u_k$. Hence, we need a representation of the sum that maintains the summation index in a symbolic form.

\subsection{Indefinite symbolic sums}

Let $A_1, A_2, A_3, \ldots,A_n$ be arbitrary sets and let $f(\mathbf a)$ be arbitrary function of $\mathbf a = [a_1, a_2, a_3,\ldots,a_n] \in A_1\times A_2\times A_3\times \ldots\times A_n$. Let us define the following sum:

\begin{equation}
\label{eq:sumfa}
\displaystyle \sum_{\mathbf a}f(\mathbf a):=
\sum_{\mathclap{a_1\in A_1, \ldots, a_n\in A_n}}f(a_1, \ldots,a_n)=
\sum_{a_1\in A_1}\sum_{a_2\in A_2}\ldots \sum_{a_n\in A_n}f(a_1, \ldots,a_n)
\end{equation}
We do not discuss the existence of \eqref{eq:sumfa} and we assume that the information about the form of $A_k$ does not matter. Hereafter, all formulae are considered as {\it symbolic expressions}. Let us introduce the following {\it indefinite symbolic sum} as a representation of \eqref{eq:sumfa}:
\begin{equation}
\label{eq:sumfaSymb}
sum(f(a_1, \ldots, a_n), [a_1,\dots, a_n]), 
\end{equation}
where the operands are, correspondingly,  a summand and a sequence of symbols over which the sum is taken. 
For example, using this representation, the sum 
$\sum_{{m, n=1}}^N|a_m-a_m|$
is expressed  as follows:
$$sum(|a_1-a_2|, [a_1, a_2]).$$

The key point of presented approach to computations using $sum$ is to prevent the expression, undergoing symbolic manipulations, from generating nested $sum$ subexpressions. This leads to the application of the following rules in the {\it automatic simplification} algorithms~\cite{Cohen}:
\begin{description}
  \item[Rule 1.]
$sum(f(\mathbf a)+g(\mathbf a),\mathbf a) \rightarrow sum(f(\mathbf a),\mathbf a)+sum(g(\mathbf a),\mathbf a)$
  \item[Rule 2.]
$f(\mathbf a) sum(g(\mathbf b),\mathbf b) \rightarrow sum(f(\mathbf a) g(\mathbf b),\mathbf b)$
  \item[Rule 3.] 
  If $\mathbf a  \cap \mathbf b=\emptyset$, then\\
\phantom{}\hspace{10mm}$sum(sum(f(\mathbf a,\mathbf b), \mathbf b), \mathbf a)\rightarrow  sum(f(\mathbf a, \mathbf b), [\mathbf a, \mathbf b]),$\\
where $[\mathbf a, \mathbf b]=[a_1,\ldots,a_n, b_1,\ldots,b_n,]$.
  \item[Rule 4.] 
$sum(0,\mathbf a) \rightarrow 0$
\end{description}

\subsection{Symbolic representation of successive approximations}

The following system forms the generalized form of \eqref{eq:3D37}:
\begin{equation}
\label{eq:genSysEqs}
u_k(\mathbf x)= c_k + f(\mathbf x) + \sum_{m \neq k} g(\mathbf x, a_m) (u_m\circ s)(\mathbf x, a_m), \;  
\;k=1,2, \ldots, n  ,
\end{equation}
where $u_k(\mathbf x)$ ($k=1,2, \ldots, n$), $c_k$ are unknown and $\circ$ denotes function composition.  The generalized form of the system can be applied in similar problems, for instance, in the effective conductivity of 2D composite materials \cite[eq.~(2.3.85)]{{GMN}}. The system \eqref{eq:genSysEqs} can be rewritten in a symbolic form:
\begin{equation*}
u_k(\mathbf x)=c_k + f(\mathbf x) + sum\left(g(\mathbf x, a_m) (u_m\circ s)(\mathbf x, a_m), [a_m] \right) , \;  
\;k=1,2, \ldots, n ,
\end{equation*}
assuming that summation excludes $a_k$. The application of the successive approximations yields following relations:
\begin{align*}
u_{k,0}(\mathbf x)&= c_{k, 0} + f_{0}(\mathbf x)\\ 
u_{k,q}(\mathbf x)&= c_{k,q} + f_{q}(\mathbf x) + sum(g(\mathbf x, a_{m,q-1}) (u_{m,q-1}\circ s)(\mathbf x, a_{m,q-1}), [a_{q-1}])\\
& k=1,2, \ldots, n,
\end{align*}
where $u_{k,q}$ denotes the approximation of $u_k$ in $q$th iteration. Moreover, the remaining expressions indexed by $q$ correspond to expressions introduced to the solution by the $q$th iteration.
For the sake of implementation, one can omit indexes $k$ and $m$. This yields relations:
\begin{align} 
\label{eq:symSysEqs1}
u_0(\mathbf x)&= c_0 + f_0(\mathbf x) \nonumber\\ 
u_q(\mathbf x)&= c_q + f_q(\mathbf x) \\
& + sum(Algebraic\_expand(g(\mathbf x, a_{q-1}) (u_{q-1}\circ s)(\mathbf x, a_{q-1})), [a_{q-1}]) \nonumber.
\end{align}
It is easy to prove, by induction and applying rules~1-4, that each approximation $u_q(\mathbf x)$ can be rewritten in a form of a sum of non-nested $sum$ expressions and the term $f(\mathbf x) + c_q$. Note that, the $Algebraic\_expand$ procedure in \eqref{eq:symSysEqs1} keeps the summand expanded and let the rules work properly. The relation \eqref{eq:symSysEqs1}, as well as the rules~1-4 can be implemented in a Computer Algebra System (CAS). One can revert the symbolic expression $u_q(\mathbf x)$ to an analytic formula, through two simple steps:
\begin{itemize}
  \item symbols $c_q$, $a_q$ corresponds to $c_k$, $a_k$ in \eqref{eq:genSysEqs};
  \item symbols $c_m$, $a_m$ for $m<q$ transform to the summation, e.g.  
  $\displaystyle
sum(h(c_3, c_4, a_3, a_4), [a_3, a_4]) \;\text{corresponds to} \; \sum_{\substack{m \neq k \\ s\neq m}}h(c_s, c_m, a_s, a_m).$
\end{itemize}

\subsection{Application to the effective conductivity}

Let us sketch the algorithm of computing the analytic approximation \eqref{eq:PoiDcf} of functions $u_k(\mathbf x)$ explicitly  up to  $O(r_0^{q+1})$  using presented model.
The procedure follows Sect.~\ref{chap1:subsec3D-D} and is expressed in the {\it Mathematical Pseudo-language} (MPL), described in details by Cohen~\cite{Cohen}, covering basic operations common for Computer Algebra Systems (CAS). The construction of the procedure is based on the application of following {\it transformation rules}, describing symbolic operations of substituting terms of a given form:
\begin{description}
  \item[Rule 5.]
$sum(f(\mathbf a),\mathbf a) \rightarrow sum(Algebraic\_expand(f(\mathbf a)),\mathbf a)$
  \item[Rule 6.]
$sum(f(\mathbf a),\mathbf a) \rightarrow sum(Series(f(\mathbf a), r_0, 0, q),\mathbf a)$
\end{description}
The $Series$ operator in the rule~6 produces a power series expansion for $f(\mathbf a)$ with respect to $r_0 $ about the point 0 to order $q$. Some CAS have the capability of implementation of transformation rules through the {\it rule-based programming}. Hence, the following procedure can be directly realized:
\begin{equation*}
  \label{eq:mpl_proc_1}
  \begin{array}{lll}
\mbox{\textbf{Procedure}} \;u(q): \\
\;\;1 \quad expr:= u_q(\mathbf x) \text{ \quad\quad // computed via \eqref{eq:symSysEqs1} }\\
\;\;2 \quad expr := Substitute(expr, \text{Rule~6})\\
\;\;3 \quad zexpr := expr \\
\;\;4 \quad zexpr := Substitute(zexpr, \mathbf x \rightarrow \mathbf a_q)\\
\;\;5 \quad \mbox{\textbf{for each }} m\leq q:\\ 
\;\;6 \quad\quad\quad zexpr := Substitute(zexpr,  c_m  \rightarrow  a_{m_j} - z(m)) \\ 
\;\;7 \quad zexpr := Solve(zexpr=0, z_q) \text{\quad\quad// based on \eqref{eq:3D212d}} \\
\;\;8 \quad\mbox{\textbf{while }} not\; Free\_of(zexpr, z):\\
\;\;9 \quad\quad\quad\mbox{\textbf{for each }}  m<q:\\ 
10   \quad\quad\quad\quad\quad  zexpr := Substitute(zexpr, z_m  \rightarrow  Reindex(zexpr, q, m))\\ 
11   \quad\quad\quad zexpr := Sequential\_substitute(zexpr, [\text{Rule~5}, \text{Rule~6}])\\
12   \quad \mbox{\textbf{for each}} \quad m:\\
13   \quad\quad\quad  expr := Substitute(expr, c_m \rightarrow  a_{m_j}-Reindex(zexpr, q, m))\\
14   \quad expr := Sequential\_substitute(expr, [\text{Rule~5}, \text{Rule~6}])\\
15	\quad Return(expr)\\
\end{array}
\end{equation*}
Lines 8-11 find constants $c_m$ by calculating  $z_q$ explicitly up to $O(r_0^{q+1})$ by successive approximations. Then, $c_m$ are substituted to the solution in lines 12-13. The $Reindex(expr, q, m)$ operator substitutes all symbols of the form $z_m, a_m$ in $expr$ by $z_{2m-q}, a_{2m-q}$ respectively, in order to prevent indices from duplicating in the resulting expression. Note that negative indices may appear, however it does not matter in the applied model of calculations. More details of the procedure $u(q)$, as well as an example implementation will be published in a separate paper.

\end{document}